\documentclass[preprintnumbers,11pt,a4paper,oneside]{article}
\pdfoutput=1

\usepackage{jheppub}
\usepackage{color}
\usepackage{graphicx}
\usepackage{wrapfig,enumerate,slashed}

\usepackage{footmisc}
\usepackage{amsmath}
\usepackage{wasysym} 
\usepackage{graphicx}
\usepackage{subcaption}
\usepackage{color}
\usepackage{comment}
\usepackage{appendix}
\usepackage{slashed}

\newcommand{\be}{\begin{eqnarray}}
\newcommand{\ee}{\end{eqnarray}}

\newcommand{\bee}{\begin{eqnarray}}
\newcommand{\eee}{\end{eqnarray}}
\newcommand{\beeq}{\begin{equation}}
\newcommand{\eeeq}{\end{equation}}

\usepackage{tikz}

\makeatletter
\gdef\@fpheader{}
\makeatother
\begin{document}

\title{Quantum Pathways for Charged Track Finding in High-Energy Collisions}

\author[a]{Christopher Brown,}
\author[b]{Michael Spannowsky,}
\author[a]{Alexander Tapper,}
\author[a,b]{Simon Williams,}
\author[a]{Ioannis Xiotidis }

\affiliation[a]{High Energy Physics Group, Blackett Laboratory, Imperial College, Prince Consort Road, London, SW7 2AZ, United Kingdom}
\affiliation[b]{Institute for Particle Physics Phenomenology, Department of Physics, Durham University, Durham DH1 3LE, U.K.}

\abstract{In high-energy particle collisions, charged track finding is a complex yet crucial endeavour. We propose a quantum algorithm, specifically quantum template matching, to enhance the accuracy and efficiency of track finding. Abstracting the Quantum Amplitude Amplification routine by introducing a data register, and utilising a novel oracle construction, allows data to be parsed to the circuit and matched with a hit-pattern template, without prior knowledge of the input data. Furthermore, we address the challenges posed by missing hit data, demonstrating the ability of the quantum template matching algorithm to successfully identify charged-particle tracks from hit patterns with missing hits. Our findings therefore propose quantum methodologies tailored for real-world applications and underline the potential of quantum computing in collider physics.}

\preprintA{IPPP/23/61}
\maketitle


\section{Introduction}

In collider physics, the endeavour of accurately associating the multitude of hits in the detectors recorded during high-energy particle collisions with the original charged particle tracks that traversed the detector emerges as a particularly challenging combinatorial problem~\cite{CMS:2014pgm, Cerati:2015vna}. The precise assignment of these detector hits is pivotal for deducing the underlying nature and dynamics that catalysed the fundamental interactions being probed in such collisions. The critical endeavour of tracking therefore fosters a deeper understanding and elucidation of new physics phenomena, thereby acting as a linchpin in advancing high-energy physics.

The gamut of issues encountered in high-energy physics often resembles database search algorithms, where the solution to a particular problem is embodied as a notable element within a specified dataset. A prime exemplification of this is identifying charged particle tracks within a detector experiment, as seen in the eminent CMS~\cite{CMS:2008xjf} and ATLAS~\cite{ATLAS:2008xda} experiments at CERN. This task can be conceptualised as a variant of a search algorithm known as template matching. The primary objective is to discern charged particle tracks traversing the tracker detector by juxtaposing the raw detector response against a pre-established database encompassing hit patterns that correspond to physical particle tracks obtained from simulation. Upon the recognition of a physical track within the data, attributes of the track, such as momentum and angular distribution, can be gleaned from the template database. This procedure is prominently recognised as \textit{Associative Memory}, and has been shown to be a highly effective approach to track finding in high-energy experiments, employing Application Specific Integrated Circuits (ASIC)~\cite{DellOrso:1988gmu} to perform the template matching. The method of template based track finding is used in modern detector experiments~\cite{BUNKOWSKI2019368,RNicolaidou_2010, Aad_2021} and is marked as one of the potential approaches to be used at future colliders.  

The proficiency of template matching algorithms is heavily contingent on the efficiency at which one can traverse through the template database. In an unstructured database comprising N elements, conventional search algorithms exhibit a scaling of $\mathcal{O}(N)$, necessitating, on average, $N/2$ queries to the database to pinpoint the matching element. Contemporary particle colliders witness an increase in the number of potential tracks encoded in the database, congruent with the escalating energy and luminosity of the collisions within the detectors. Concurrently, since the advent of advanced tracking technology, tracking detectors have been evolving to become highly granular, thereby amplifying the resolution of the tracks and, consequently, the quanta of track patterns necessitated to be encoded into the template database. As the frontier of high-energy and high-luminosity experiments beckons, the practice of identifying charged particle tracks via Associative Memory is confronted with a duo of challenges: (1) the rapidly increasing number of tracks encoded in the template database demands a significant amplification in storage capacity to accommodate the probable tracks, and (2) the temporal resources required to sift through a burgeoning number of tracks is inefficient for modern tracking objectives.

With its rapid and continuous development, quantum computing offers a paradigm shift in information science and has the potential to revolutionise modern computational techniques. Particle physics will benefit from any speedup that quantum computers can provide and the devices' ability to compute in a regime that has never been accessible before. Already, there has been a quickly developing research effort into proof-of-principle algorithms for applications in particle physics ranging from the simulation of quantum field theories~\cite{Davoudi:2022xmb, Fromm:2023npm, Jordan:2014tma, Kan:2021xfc, Ciavarella:2021nmj, PRXQuantum.2.030334, Kane:2022ejm} and collision events~\cite{Bepari:2020xqi, Bepari:2021kwv, Gustafson:2022dsq, Bauer:2019qxa, Li:2021kcs, Barata:2023clv, Chawdhry:2023jks}, to event classification~\cite{Araz:2022haf, Blance:2020ktp} and analysis~\cite{Wu:2020cye, Mott:2017xdb}. Quantum tracking algorithms have gained a lot of interest~\cite{Shapoval_2019, Gray:2022fou, Duckett:2022ccc, Zlokapa:2019tkn, Bapst:2019llh} in an attempt to combat the problems facing classical techniques. Quantum computers offer a solution to the limitations of Associative Memory. The exponentially growing Hilbert space of qubit-based systems allows large datasets to be encoded onto quantum devices with efficient resource usage~\cite{ventura1998quantum, Shapoval_2019}. Furthermore, it has been shown that a polynomial speedup can be achieved for search algorithms by leveraging the Grover Search Algorithm~\cite{grover1996fast, PhysRevLett.79.325}, which has been suggested as a tool to achieve the crucial speedup required for Associative Memory to be effective for tracking algorithms~\cite{Shapoval_2019}.

This paper proposes a proof-of-principle quantum algorithm which extends on the regular Grover search approach to track finding via Associative Memory, proposed in Reference~\cite{Shapoval_2019}, by abstracting the oracle operation to perform a template matching algorithm to match detector-hit data with a pre-established database of physical tracks. Following the oracle construction method of Reference~\cite{PhysRevResearch.4.023006}, it will be shown that a single, general oracle operation can be constructed for the template matching approach to successfully identify particle tracks. Additionally, we will demonstrate that the template matching approach further improves on the regular Grover search approach by allowing for data with missing hits to be efficiently reconstructed, a highly non-trivial task for classical tracking algorithms.


\section{Grover Search and Quantum Amplitude Amplification}\label{sec:Grover}

The Grover Search is an optimal quantum search algorithm~\cite{grover1996fast, PhysRevLett.79.325} which amplifies the amplitudes of \textit{marked} states within a uniformly distributed database to successfully identify elements of interest, achieving a polynomial speedup over classical search techniques for unstructured databases. Consider an unstructured dataset of $N$ elements $X=\{x_1, x_2, \hdots, x_N\}$ with one or more elements of interest, $m_j$, which can be encoded onto $n=\log_2(N)$ qubits as an equal superposition, 

\begin{equation}\label{eqn:equalState}
\vert s \rangle = \mathcal{A}_G \, \vert 0 \rangle^{\otimes n} =  \frac{1}{\sqrt{N}} \sum_i \vert x_i \rangle,
\end{equation} 
where $\mathcal{A}_G = H^{\otimes n}$ is an $n$-qubit Hadamard transformation which prepares the state $\vert s \rangle$, and the states $\vert x_i\rangle$ encode the elements $x_i$ in the computational basis on the quantum device. The Grover Search aims to identify the elements of interest, $m_j$, in the database $X$ and amplify their amplitudes. To first identify the marked elements, one can define a Boolean function, $f(x)$, such that

\begin{equation}
f(x) = \begin{cases} 1 & \textrm{if } x = m, \\ 0 & \textrm{otherwise.} \end{cases}
\end{equation} 
This function can then be used to construct the \textit{oracle},

\begin{equation}
S_f \vert x \rangle = (-1)^{f(x)} \vert x \rangle,
\end{equation} 
such that the amplitude of an element of interest is \textit{marked} by inverting the amplitude of the state and leaving all other states unchanged. Marking the states alone is not enough to successfully identify the elements of interest, as a measurement at this stage will still return each element with equal probability. Therefore, one must amplify the amplitudes of the marked states such that a measurement returns one of the marked states with a high probability. Geometrically, the amplification process can be modelled as a reflection of the whole system about the equal state $\vert s \rangle$ from Equation~\ref{eqn:equalState}, reducing the amplitudes of the unmarked states, and amplifying the marked states. This can be achieved by applying the \textit{Grover diffuser}, which has the form 

\begin{equation}\label{eqn:diffuser}
D = \mathcal{A}_G^\dagger S_0 \mathcal{A}_G,
\end{equation}
where $S_0$ is a phase inversion on the zero state, and in the case of the Grover Search, $\mathcal{A}_G^\dagger =\mathcal{A}_G$, as the Hadamard transform is Hermitian. 

Combining the diffuser with the oracle, one step of the algorithm can be defined as a single, unitary operation, the \textit{Grover Iterator},

\begin{equation}\label{eqn:GroverIterator}
\mathcal{Q} = \mathcal{A}_G^\dagger S_0 \mathcal{A}_G S_f,
\end{equation}
which can be applied iteratively to amplify the amplitudes of all states of interest in the database. For an unstructured database of $N$-elements with $m$-elements of interest, the optimal number of applications of $\mathcal{Q}$ to achieve the highest probability of measuring a state of interest is 

\begin{equation}\label{eqn:iterations}
t = \Bigg \lfloor \frac{\pi}{4} \sqrt{\frac{N}{m}} \Bigg \rfloor.
\end{equation}
The Grover Search, therefore, scales as $\mathcal{O}(\sqrt{N})$, providing a remarkable polynomial speedup over a classical search algorithm. Consequently, the Grover Search offers a substantial speedup when searching large databases, typical of those produced by modern particle physics experiments. 

It should be noted that it is possible to construct a database for which the number of elements of interest, $m$, is not known \textit{a priori}. Therefore it is not clear how many iterations of $\mathcal{Q}$ should be applied to reliably return an element of interest from the database upon measurement. To establish $m$, one can use Quantum Counting~\cite{Brassard:1998vj} which leverages Quantum Phase Estimation~\cite{Kitaev:1995qy} to estimate the number of interesting states in the database, and, by extension, the number of applications of $\mathcal{Q}$. For the examples considered in this paper, the number of elements of interest is known by construction of the database, however future implementations may benefit from the Quantum Counting routine. 


\subsection{Quantum Amplitude Amplification}

The Grover algorithm performs a search on a uniform, unstructured database encoded onto $n$-qubits using a Hadamard transformation, $\mathcal{A}_G = H^{\otimes n}$. However, it is often the case that it is not efficient to encode the database as a uniform superposition, but instead as an arbitrary state, $\vert s^\prime\rangle$, prepared using the unitary operation $\mathcal{A}$. Quantum Amplitude Amplification (QAA)~\cite{Brassard_2002} is a generalisation of the Grover Search algorithm which can perform a search on $\vert s^\prime\rangle$ by modifying the Grover Iterator from Equation~\ref{eqn:GroverIterator}. 

\begin{figure}[t]
\centering
\includegraphics[width=0.9\textwidth]{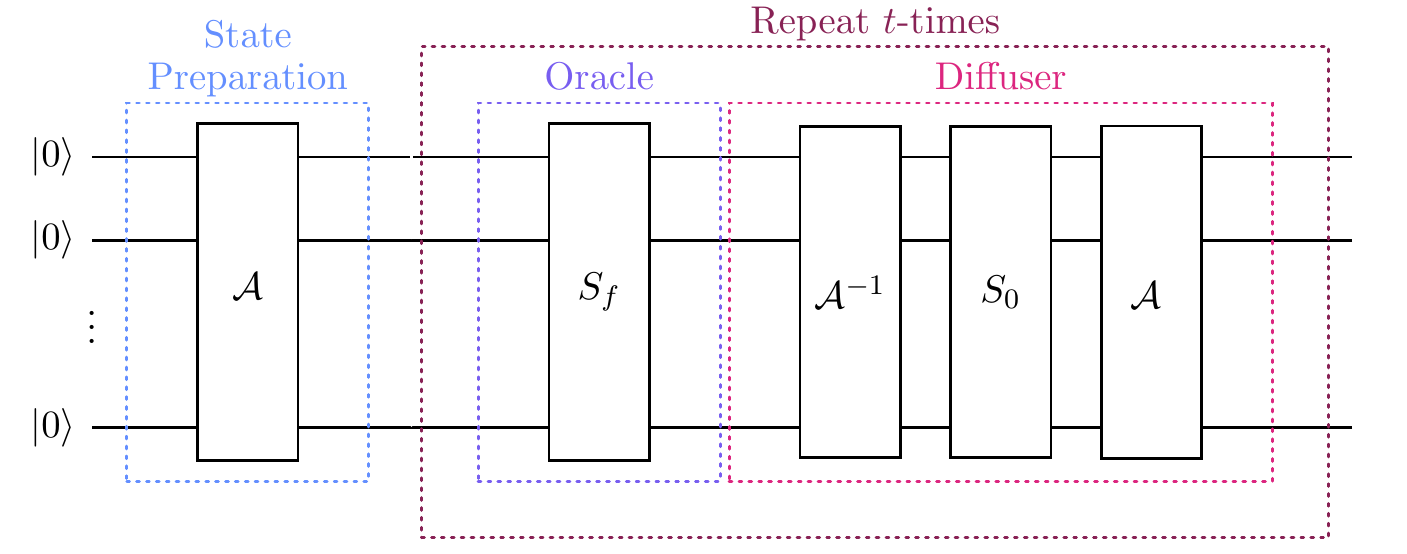}
\caption{Schematic circuit diagram for Quantum Amplitude Amplification (QAA) on $n$ qubits. The circuit is initialised by preparing an arbitrary state using the unitary operation $\mathcal{A}$. The state is then parsed to the QAA routine which amplifies the amplitudes of interesting states in the initial state. The QAA routine is applied $t$-times to return an interesting state with high probability, upon measurement. The QAA routine is constructed from two operations: the oracle, which marks the interesting states by inverting their phase, and the diffuser, which performs a reflection to amplify the amplitudes of the marked states. }\label{fig:QAACircuit}
\end{figure}

As shown in Equation~\ref{eqn:diffuser}, the amplification of a marked state is performed by reflecting around the state $\vert s \rangle$. Generalising to an arbitrary initial state, the diffuser operation now reflects around the state $\vert s^\prime\rangle$, and thus has the form $D^\prime = \mathcal{A} S_0 \mathcal{A}^\dagger$. The Grover Iterator becomes,

\begin{equation}\label{eqn:QAAit}
\mathcal{Q} = \mathcal{A} S_0 \mathcal{A}^\dagger S_f,
\end{equation}
such that the Grover Iterator from Equation~\ref{eqn:GroverIterator} can be retrieved by identifying the preparation of the state $\vert s \rangle$ as an $n$-qubit Hadamard transform. Figure~\ref{fig:QAACircuit} shows a schematic circuit diagram for Quantum Amplitude Amplification. 

\subsection{Oracle construction}

The explicit form of the oracle, $S_f$, has so far remained an undefined black box in both the Grover and QAA routines. The only speculation is that the oracle must mark any interesting states within the database by inverting the phase of the marked states' amplitudes. Consider the example where a database of four states is encoded onto two qubits via a Hadamard transform, 

\begin{equation}
H^{\otimes 2} \vert 0 \rangle ^{\otimes 2} = \frac{1}{\sqrt{4}} \sum_{i=0}^{3} \vert i \rangle.
\end{equation}
It is possible to define an oracle which will search this database for the state $\vert 11 \rangle$ by applying a controlled-$Z$ gate operation, which will apply the $Z$-gate operation to the target qubit if the control qubit is in the `1' state. The oracle operation therefore has the form

\begin{equation}
S_f : \mathbb{I} \otimes \vert 0 \rangle \langle 0 \vert + Z \otimes \vert 1 \rangle \langle 1 \vert.
\end{equation}
Acting the oracle on the initial state, we find 

\begin{equation}
\frac{1}{\sqrt{4}} \sum_{i=0}^{3} \vert i \rangle = \frac{1}{\sqrt{4}} \Big[ \vert 00\rangle + \vert 10\rangle + \vert 01\rangle - \vert 11 \rangle \Big],
\end{equation}
thus the state $\vert 11 \rangle$ has been marked by the oracle. However, if the target state now changes from $\vert 11 \rangle$, the form of the oracle has to change. For the problem of track finding, this is limiting as one would have to transpile the circuit each time a data string is retrieved from the detector to correctly search for, and identify, a matching hit-pattern template in the database. In Section~\ref{sec:qTrack}, an algorithm is proposed that removes this limitation by generalising the oracle construction, allowing for the same circuit to be used for all data retrieved from the detector, without having to know how to construct the oracle \textit{a priori}. 


\section{Track Finding via Associative Memory}

\begin{figure}[t]
\centering
\includegraphics[width=0.6\textwidth]{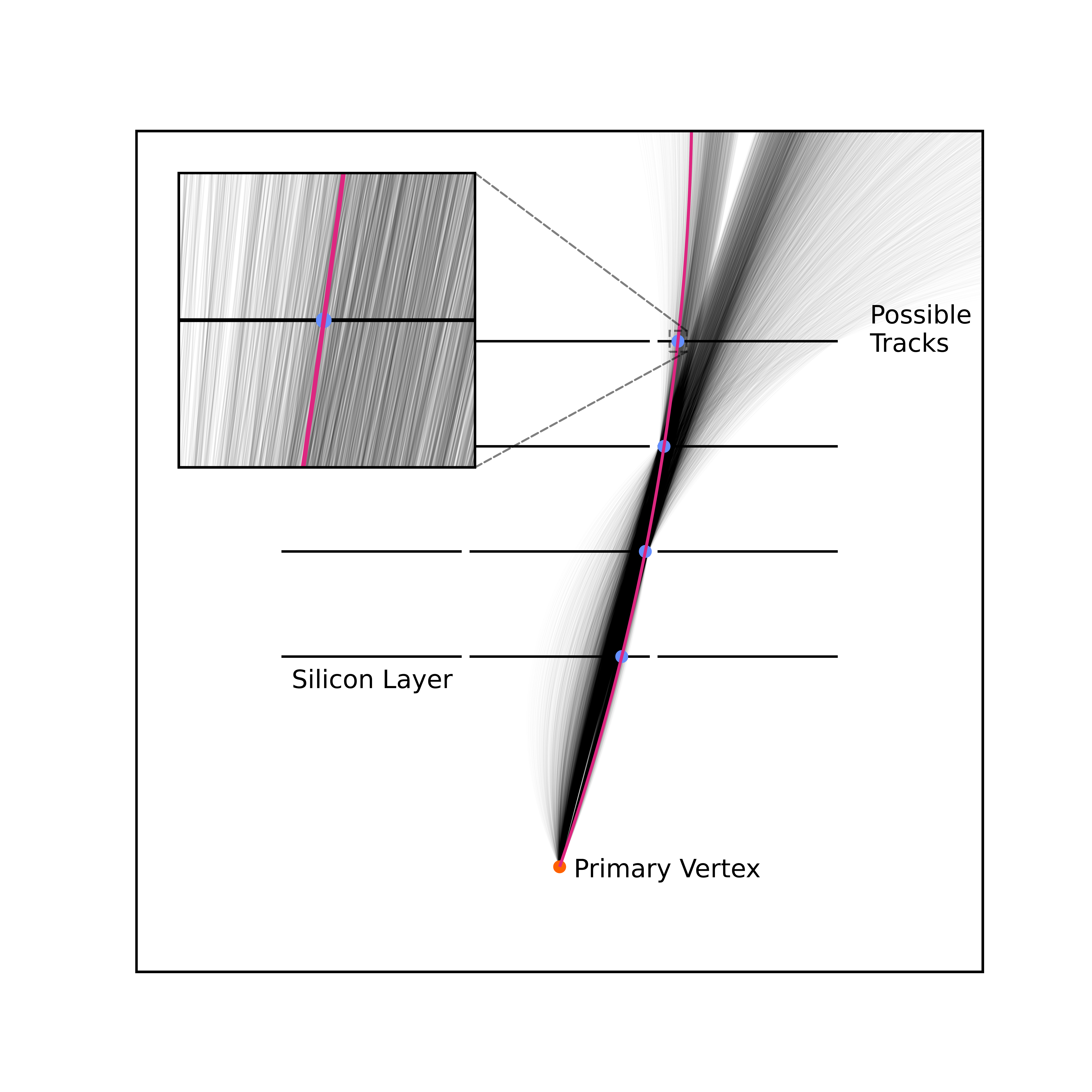}
\caption{A single track through a 12-module detector, arranged in four layers of three detector modules. The red track shows the ``true" track through the detector, with the blue circles representing the hits in the detector. The black tracks show a selection of possible tracks which can also lead to this hit pattern in the detector. Increasing the granularity of the detector decreases the number of tracks corresponding to a single hit pattern, but increases the combinatorial challenge of finding possible hit patterns left by charged particles. }\label{fig:Tracker}
\end{figure}

Modern high-energy collider experiments collide particles together at unprecedented energies in the centre of close-to-fully hermetic detectors. These detectors comprise many sub-detector regions immersed in strong magnetic fields. The experiment aims to precisely reconstruct the energy and momentum of each particle created in the collision event to unveil the underlying physics in play. The reconstruction of a high-energy collision event can be separated into three main steps: (1) the reconstruction of the charged particle trajectories as they traverse the detector layers through particle tracking, (2) the determination of the particle energies using calorimetry, and (3) the reconstruction of muons in dedicated tracking modules on the outer layer of the detector device. From this process, essential characteristics of the underlying physics can be obtained. For example, the particle species can be identified, and any missing energy can be established. This paper will focus on the first step, designing a quantum algorithm to identify charged particle tracks in the detector efficiently. 

To successfully record the trajectory of a charged particle within a tracking detector, a method for measuring the particle's position without disturbing its path is required. In state-of-the-art collider experiments such as the CMS~\cite{CMS:2008xjf} and ATLAS~\cite{ATLAS:2008xda} experiments on the Large Hadron Collider at CERN, sub-millimetre-thick layers of silicon sample a particle's trajectory by recording a hit every time the particle passes through a layer. By applying strong magnetic fields the trajectory of a charged particle can be curved proportionally to the inverse of the particle's momentum. Thus, with a position granularity of tens of micro-metres, the tracking detector can accurately reconstruct the particle's trajectory, allowing for the particle's charge and momentum to be inferred from the track's curvature. Furthermore, the high precision of silicon trackers allows for the accurate reconstruction of jets, cones of high-energy hadrons emerging from the hadronisation of colour-charged particles resulting from the high-energy collision. Displacements of hundreds of micro-metres can be reconstructed, allowing for the individual constituents of the jet to be separately resolved. As a result, the detectors can identify complete decay chains emerging from the high-energy collision event.

In the tracking detectors, the only information collected is the position at which each track traversed a silicon layer. The hits belonging to an individual particle track must, therefore, be identified from the raw detector output. Once identified, a fitting operation is performed to dress the track with parameters such as the azimuthal angle, $\phi$, at which the particle has been produced, and the reconstructed transverse momentum, $p_T$. Due to the wide range of possible interactions, the paths of the charged particles in the tracking detectors vary, and the possible combination of hit patterns they produce is extensive. Furthermore, each hit pattern is associated with many tracks. For example, Figure~\ref{fig:Tracker} indicates some of the possible tracks that have the same hit pattern in a simple, 12-module detector. With increasingly granular detectors, the number of possible hit patterns is increasing to unmanageable levels. For a single particle crossing the detector, identifying the hits corresponding to the particle's track is seemingly a trivial task. However, in particle collisions, thousands of charged particles traverse the detector every fraction of a second. Each particle leaves a set of hits in the detector, leading to tens of thousands of hits in the detector from the particle trajectories, all overlapping. Therefore, the reconstruction of particles becomes a highly challenging combinatorial problem. Reference~\cite{Cerati:2015vna} highlights the increasing challenge of track reconstruction at high pile-up and presents analysis of reconstruction times for current track reconstruction methods.

Classical techniques such as Associative Memory, which employs a template matching approach to track finding, have been shown to be highly effective at identifying hit patterns in the tracking detectors~\cite{Bardi:1997hm}. However, as the number of particles through the detector increases with the collider energy and luminosity, and tracking detectors become more granular, the number of hit patterns that need to be stored and compared becomes increasingly unmanageable, and the time taken to find the correct match grows quickly. With the exponentially growing Hilbert space of a qubit system and the polynomial speedup of Quantum Amplitude Amplification (QAA), quantum computers provide a potentially powerful tool for tackling the track finding problem. In Section~\ref{sec:qTrack}, a proof-of-principle quantum tracking algorithm is proposed, which harnesses the advantage over the QAA routine. In Section~\ref{sec:missingHits}, it will be shown that this algorithm can be extended to handle imperfect data which has missing hits efficiently. 


\section{Quantum Template Matching for Track Finding}\label{sec:qTrack}

To successfully identify tracks in hit data from the detector via a quantum template matching algorithm, the oracle must have a general construction to identify the correct track without prior knowledge of the input data. Following the oracle construction from Reference~\cite{PhysRevResearch.4.023006}, it is possible to design a general oracle, $\tilde{S}_f$, by introducing an additional register to the QAA circuit in Figure~\ref{fig:QAACircuit}, the \textit{data} register. The retrieved detector-hits data from the experiment is encoded on this register for each event with the unitary operation $\mathcal{A}_D$. The register retained from the QAA routine will now encode the template database, the \textit{template} register. For a tracker in the same configuration as Figure~\ref{fig:Tracker}, with 12 tracker modules arranged in layers of threes, there are 15 possible hit patterns for particles traversing the detector, neglecting multiple track signatures and requiring one hit per detector layer. These hit patterns are one-hot encoded into bit strings of 12 bits, with each bit corresponding to a detector module. If a hit is detected on the module, the bit is flipped to the `1' state. Otherwise, it remains in the `0' state\footnote{The choice of one-hot encoding has been made to allow for the algorithm to deal well with perfect and imperfect data from the detector, as will be outlined in Section~\ref{sec:missingHits}. Other choices of encoding may prove to be more optimal, for example encoding each track in the computational basis.}. The templates are encoded onto the template register as a linear superposition of all possible tracks through the unitary operation $\mathcal{A}_T$. The individual hit-pattern encodings are displayed in Table~\ref{tab:tracks}. The state preparation has the general form

\begin{equation}
\mathcal{A}_D\vert 0 \rangle^{\otimes n} \otimes \mathcal{A}_T \vert 0 \rangle^{\otimes n},
\end{equation}
where $n=12$ for the example from Figure~\ref{fig:Tracker}. Ideally, the database of hit-pattern templates would be loaded onto the device from Quantum Random-Access Memory (QRAM)~\cite{PhysRevLett.100.160501}, however limitations on the ability to realise QRAM mean that, currently, the database must be prepared via state preparation. State preparation is a highly non-trivial task, and has been shown to necessitate exponential circuit depths to construct an arbitrary quantum state~\cite{Xiaoming10044235}. By leveraging ancillary qubits, this scaling can be reduced to polynomial scaling in circuit depth, though at the potential cost of an exponentially growing number of ancillary qubits~\cite{PhysRevA.83.032302, PhysRevResearch.3.043200, PhysRevLett.129.230504, rosenthal2023query}. For the algorithm proposed here, the state preparation routine from Qiskit~\cite{Qiskit}, which employs a recursive initialisation algorithm with optimisation~\cite{Shende_2006}, has been used to load the one-hot encoded track templates onto the device.

\begin{table}[t]
\center
\begin{tabular}{c|c|c|c|c|c}
Track 	&  Encoding 		&  Track 	&  Encoding 		& Track	& Encoding 		\\ \cline{1-6}
1		& 010010010010	&	6	& 010010001001	&	11	& 001001001010 	\\ 
2		& 010001001001	&	7	& 001001001001	&	12	& 010100100010	\\
3		& 010100100100	&	8	& 010010010100	&	13	& 100100010010	\\
4		& 100100100100	&	9	& 010010100100	&	14	& 100100100010	\\
5		& 010010010001	&	10	& 001001010010	&	15	& 010001001010
\end{tabular}
\caption{Full list of track templates for all possible hit-patterns in a 12 module detector, with the modules arranged in four layers of three modules. Each track is required to have one hit in each layer. The track templates are one-hot encoded, with each bit corresponding to a detector module. If a hit has been detected in the module, the bit is flipped to the `1' state, otherwise it remains in the `0' state. The bit strings read left to right, with the first three bits corresponding to the first detector layer, the next three bits corresponding to the second layer, and so on.}\label{tab:tracks}
\end{table}

\begin{figure}[t]
\centering
\includegraphics[width=0.9\textwidth]{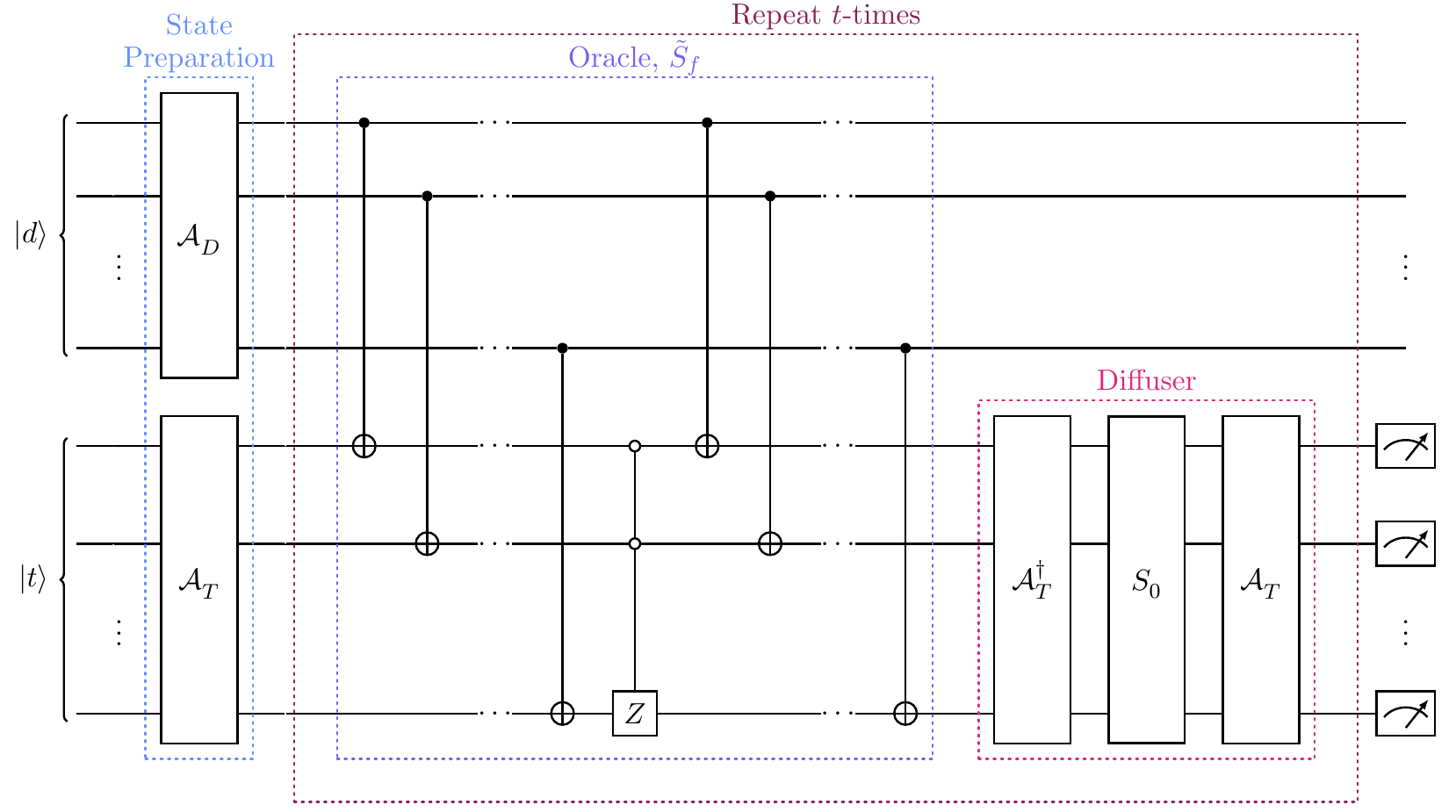}
\caption{Schematic circuit diagram for the quantum template matching algorithm for charged track finding. The circuit comprises two registers, the data register, $\vert d \rangle$, and the template register, $\vert t \rangle$. The state preparation step encodes hit data from the detector onto the data register, and the database of hit-pattern templates onto the template register using the unitary operations $\mathcal{A}_D$ and $\mathcal{A}_T$ respectively. The Quantum Amplitude Amplification (QAA) routine is then applied $t$-times to correctly identify the hit pattern within the database, with high probability. The general oracle marks the state in the template database which corresponds to the hit pattern from the detector, encoded on the data register. The diffuser operator then amplifies the marked amplitudes. A measurement is then performed to return the matched template.}\label{fig:templateMatchingCircuit}
\end{figure}

To construct the general oracle, $\tilde{S}_f$, we allow for the oracle to now act across the two registers, controlling from the data register and applying a series of \textsc{cnot} operations to the template register. If the hit pattern encoded onto the data register is in the template database, then the corresponding state on the template register will be flipped to the zeroth state. The matched state can be marked by applying a phase inversion on the zero state, $S_0$. Finally, the oracle returns the marked state to its original bit combination by applying the series of \textsc{cnot} operations again, in the same order, controlling from the data register and acting on the template register. Through this oracle operation, $\tilde{S}_f$, the track-template matching the hit-pattern encoded on the data register is marked with a negative phase, without the need for a bespoke oracle operation designed for the specific input data. To amplify the marked state, the QAA diffuser from Equation~\ref{eqn:QAAit} is then applied to the template register, where here $\mathcal{A}=\mathcal{A}_T$. To achieve the greatest probability of selecting the correct track, the oracle and diffuser are then applied $t$-times, according to Equation~\ref{eqn:iterations}. Figure~\ref{fig:templateMatchingCircuit} shows a schematic of the circuit for the quantum template matching algorithm, outlining the structure of the oracle explicitly. 

\begin{figure}[t]
\centering
\begin{subfigure}{0.49\textwidth}
\includegraphics[width=\textwidth]{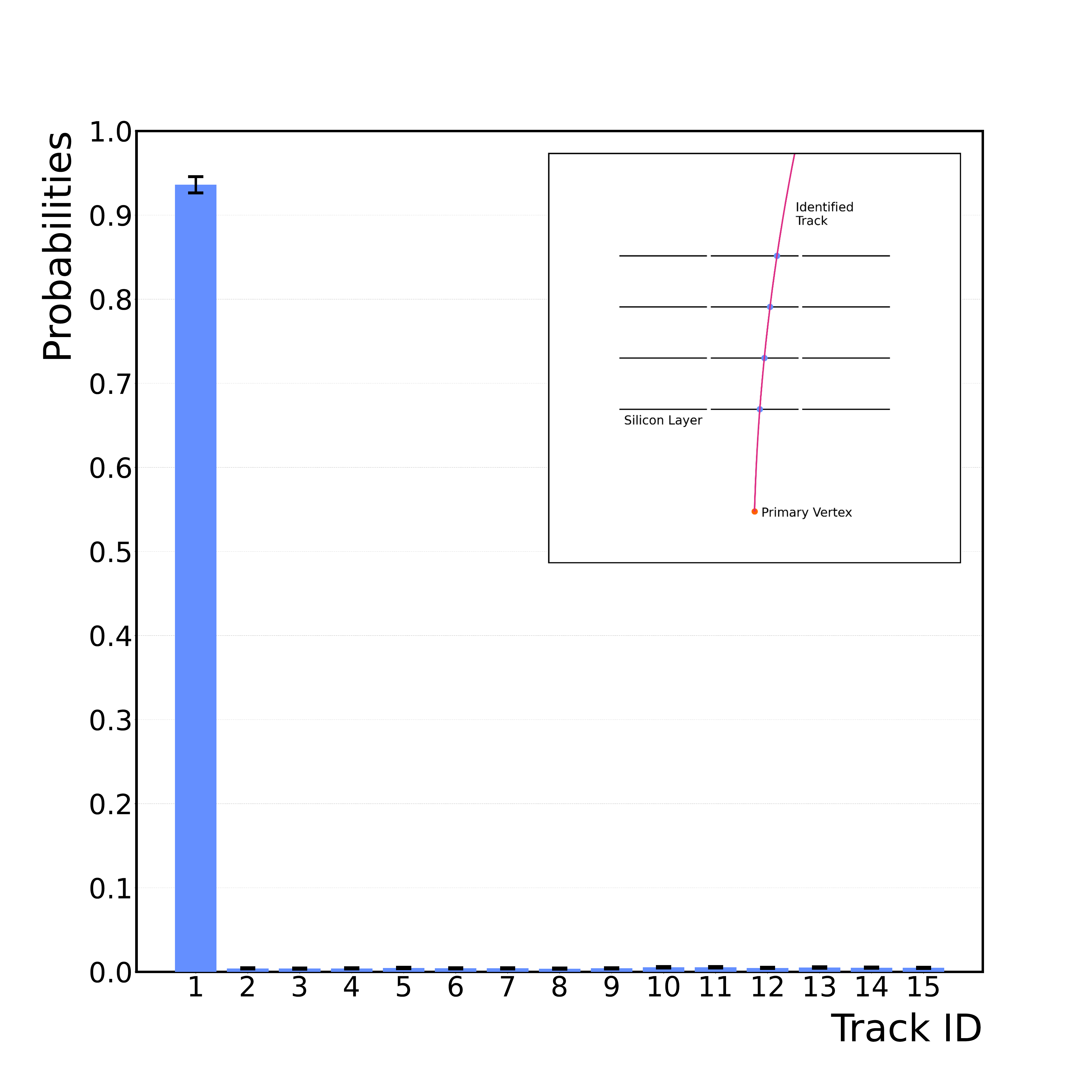}
\caption{}
\end{subfigure}
\begin{subfigure}{0.49\textwidth}
\includegraphics[width=\textwidth]{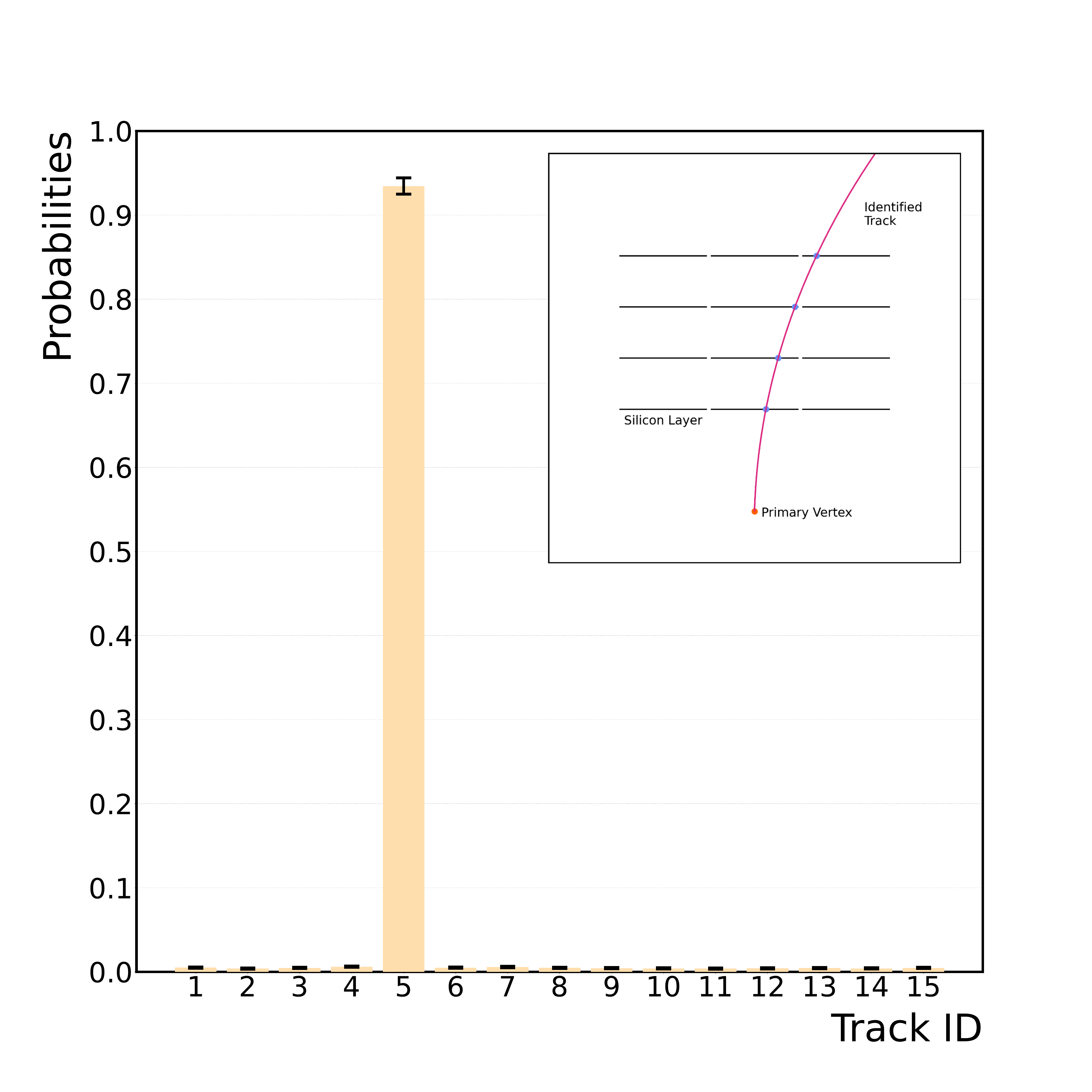}
\caption{}
\end{subfigure}
\caption{Results from the quantum template matching algorithm for two detector-hit data scenarios. Figure (a) and (b) show the correct identification of Tracks 1 and 5 from Table~\ref{tab:tracks}, shown in the top right-hand corner of the plots. The algorithm successfully identifies the correct hit-pattern templates from the database with high probability, greater than 90\%. The algorithm requires three iterations of the QAA routine, and has been run on the \texttt{qasm\_simulator} for $10^4$ shots on the device.}\label{fig:12q_match}
\end{figure}

Adopting the procedure of the quantum template matching algorithm for track finding allows for data from the detector to be parsed into the quantum algorithm and matched to a track template ``on-the-fly", as the circuit is general for all possible hit patterns handed to the algorithm. The data is one-hot encoded onto the device using the unitary operation $\mathcal{A}_D$, which applies a series of \textsc{not}-gate operations to load the data onto the device. The efficiency of the track finding algorithm has been tested for two hit patterns, corresponding to Tracks 1 and 5 in Table~\ref{tab:tracks}. To successfully determine the matching efficiency, the circuit has been run for three iterations of the QAA routine, and for $10^4$ shots on the \texttt{qasm\_simulator} without a noise model\footnote{The \texttt{qasm\_simulator}  is a 32-qubit quantum simulator that simulates a fully fault-tolerant quantum device, without noise effects.}. The results are displayed in Figure~\ref{fig:12q_match}, showing that the correct match is achieved with high probability, greater than $90\%$ efficiency.

The success in matching the data to the correct hit pattern with very high probability and the QAA routine's polynomial speedup over classical search algorithms means the algorithm is well suited to the track finding problem. Furthermore, in practice one would only have to run a small number of shots of the circuit to retrieve the correct track match with high probability, and remarkably requiring only one shot of the circuit if the track pattern is known to the be in the database. Therefore, the quantum template matching provides a fast and efficient approach to charged-particle track finding on a quantum device. However, the circuit from Figure~\ref{fig:templateMatchingCircuit} requires the data to match precisely with a hit-pattern template in the database. In practice, this is not always the case, as data from the detector may be missing hits from specific tracking modules. In Section~\ref{sec:missingHits}, it will be shown that, by modifying the oracle $\tilde{S}_f$, the quantum template matching algorithm can identify possible tracks in imperfect data, a highly non-trivial task for current classical techniques. 


\section{Track Finding on Data with Missing Hits}\label{sec:missingHits}

One of the primary challenges in track finding via Associative Memory arises when a particle passes through the detector and one or more of the detector modules on its trajectory fails to register a hit. Current, state-of-the-art track-reconstruction techniques struggle with this scenario as the combinatorics between layers with missing hits quickly become unmanageable. Overcoming this problem is paramount as the energy and luminosity of colliders are increasing, and the detectors are becoming more granular, elevating the combinatorial problem. In this Section, the quantum template matching algorithm from Section~\ref{sec:qTrack} is extended to allow for the identification of tracks from imperfect data, without an increase in computation complexity or resources.

\begin{figure}[t]
\centering
\includegraphics[width=0.9\textwidth]{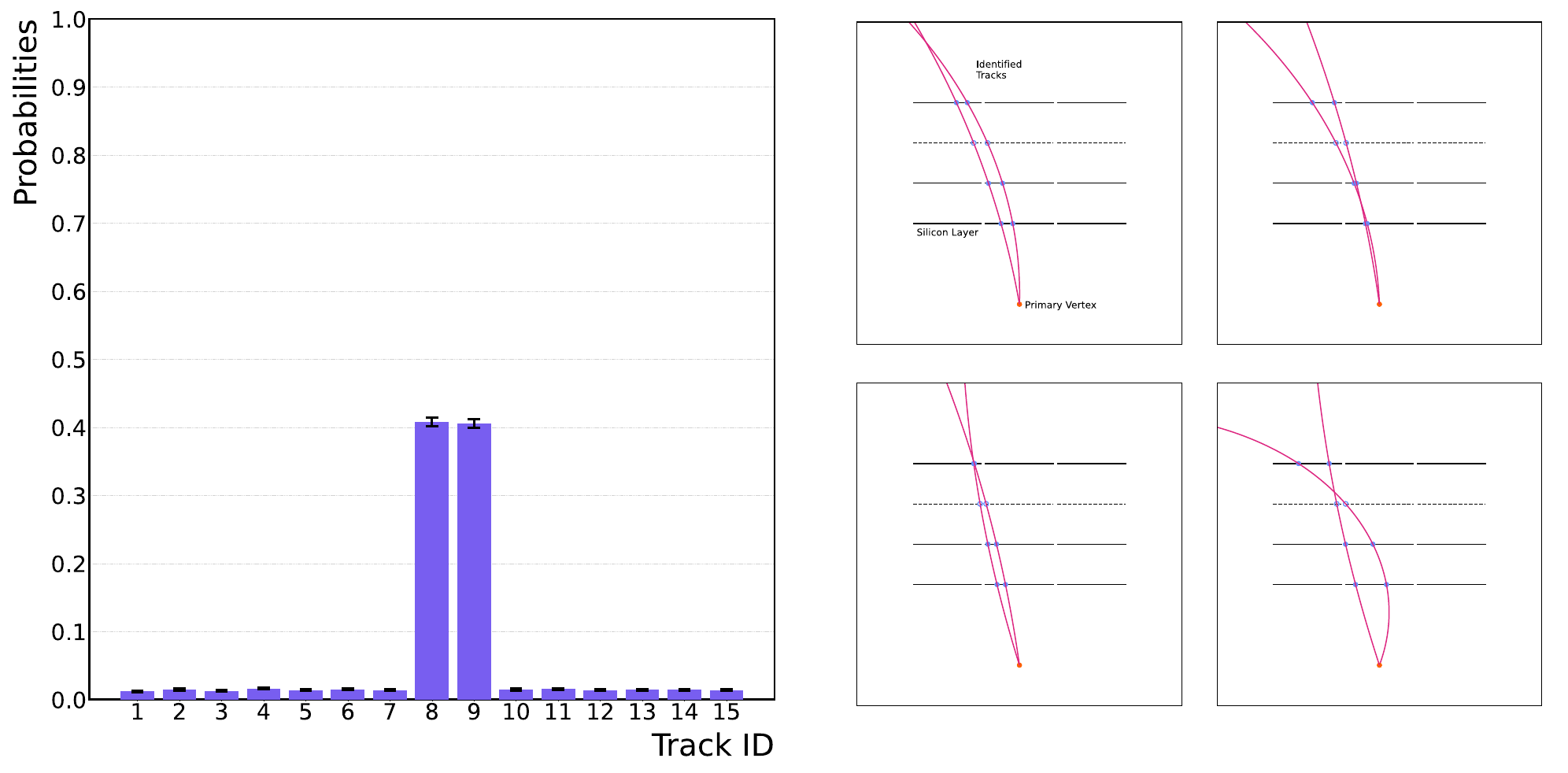}
\caption{Results from the quantum template matching algorithm with the modified oracle for data with a missing hit in the third detector layer. The results show the correct identification of the two possible hit patterns, Tracks 8 and 9 from Table~\ref{tab:tracks}. The algorithm successfully identifies the correct high-pattern templates with high probability, greater than 80\%. The algorithm requires two iterations of the QAA routine, and has been run on \texttt{qasm\_simulator} for $10^4$ shots on the device}\label{fig:missingHits}
\end{figure}

In the quantum template matching circuit shown in Figure~\ref{fig:templateMatchingCircuit}, the oracle, $\tilde{S}_f$, is essential for accurately selecting and marking the identified track by comparing, exactly, the bit strings in the data register and the template register. Consider parsing a hit pattern to the algorithm which does not contain a hit in the third layer of a detector like the one shown in Figure~\ref{fig:Tracker}. Running the algorithm for many shots would not return a decisive answer to which hit pattern matches the trajectory of the particle through the tracker, as the hit pattern without the third hit is not in the template database. To combat this problem, the oracle can be modified to correctly identify a match in the hit-pattern template database.

Using the example of a hit missing in the third detector layer, we now modify the oracle $\tilde{S}_f$ such that it does not act on the qubits corresponding to the third detector layer, and acts only on the ``good" subset of qubits corresponding to the other layers. The modified oracle, $\tilde{S}_f^{\,\prime}$, follows the same form as $\tilde{S}_f$, but acts only on the good subset of qubits: first, a series of \textsc{cnot}-operations is applied, controlling from the good subset of qubits in the data register and acting on the corresponding good subset of qubits in the template register. If there is a match, the good subset of qubits in the template register will have been flipped to the zero state. The matching state is then marked using a phase inversion on the zero state, $S_0$. Finally, the \textsc{cnot}-operations are reapplied, and the template register is returned to its original state with the exception of any matched state having a negative phase. The oracle $\tilde{S}^{\,\prime}_f$ therefore has the same effect as the oracle $\tilde{S}_f$ from Section~\ref{sec:qTrack}, but now acting only on the good subset of qubits corresponding to the tracking layers which are operating correctly. 

Employing this oracle, $\tilde{S}_f^{\,\prime}$, in the quantum template matching algorithm from Section~\ref{sec:qTrack} will return all possible hit-pattern templates with the matching good subset of qubits, allowing for efficient identification of the particle's trajectory through the detector. For the example outlined here, two states corresponding to Track 8 and 9 from Table~\ref{tab:tracks} will be marked, therefore Equation~\ref{eqn:iterations} states that two iterations of the QAA routine will yield the best match\footnote{In practice, Quantum Counting~\cite{Brassard:1998vj} can be used to determine the number of interesting states, $m$.}. Figure~\ref{fig:missingHits} shows the results from $10^4$ shots on the \texttt{qasm\_simualtor} for two iterations of the QAA routine using the modified oracle. The algorithm successfully predicts the possible path that the particle could have taken through the tracker, returning correct hit patterns for Tracks 8 and 9 from Table~\ref{tab:tracks}. Remarkably, the computational complexity and the required quantum resources do not increase when dealing with imperfect data, which is not the case using classical techniques. On the right of Figure~\ref{fig:missingHits}, an illustrative number of combinations of tracks passing through the two hit patterns shows how the combinatorics for this problem will increase dramatically for missing-hits data. 

In a modern silicon detector the efficiency of each detector module is very high and so missing hits are usually attributed to cooling or power faults, which can be quickly identified during data quality monitoring~\cite{Butz:2018dum}. Therefore, the oracle, $\tilde{S}_f^{\,\prime}$, can be easily modified to ensure efficient tracking by removing the qubits corresponding to the faulty tracker modules from the oracle operation. However, it is not always known which detector modules have failed to record a hit, thus the choice of which part of the data bit-string to examine is not clear. To effectively deal with this situation, the oracle can be modified by randomly removing \textsc{cnot}-operation controls to correctly identify possible hit-pattern matches in imperfect data to a high level of accuracy. Due to the extreme combinatorics in modern particle collider experiments, this is becoming an unmanageable problem for classical approaches, such as Associative Memory. The algorithm presented here can match both perfect and imperfect data, retrieving the correct match with high probability without any increase in computational complexity or resources for the latter. The simple but effective algorithm, therefore, provides an advantage over classical template matching techniques, both in polynomial speedup and the ability to match data with missing hits. This speedup and accuracy will become necessary as the field moves to an era of higher energies and luminosities. 

\section{Conclusion} 

Charged-track finding in high-energy particle collisions is a complex combinatorial task, fraught with challenges stemming from the sheer volume of data, noise, and intricacies of particle interactions. In this article, we present general and extendable quantum algorithms for the identification of particle tracks through a detector. As an application, a simplified detector model has been used, constructed from 12 detector modules arranged in four layers of three tracking modules. The quantum algorithms employ a novel oracle design to successfully identify particle tracks traversing the detector by matching detector-hit data to a hit-pattern template in a pre-established database of possible hit patterns. By abstracting the Quantum Amplitude Amplification (QAA) routine to encompass an additional data register, the identified template has then been amplified to deliver the correct match upon measurement. Exploiting the established polynomial speedup provided by the QAA routine~\cite{Brassard_2002}, the quantum template matching algorithm provides an advantage over classical tracking techniques via Associative Memory. Figure~\ref{fig:12q_match} contains the results from running the quantum template matching algorithm on the \texttt{qasm\_simulator} for $10^4$ shots, showing that data parsed to the algorithm has been correctly matched to a hit-pattern template. 

Confronting the prevalent issue of data with missing detector hits, the quantum template matching algorithm has been adapted and tested to mitigate the complexity of reconstructing tracks from imperfect data. By modifying the general oracle from Section~\ref{sec:qTrack}, the quantum template matching algorithm has been used to correctly identify hit-pattern templates for a track traversing the detector with one detector layer failing to register a hit. This task is highly non-trivial for classical track-identification techniques. Remarkably, the quantum resources required, and the complexity of the circuit, do not increase for imperfect data, providing a quick and efficient method for identifying tracks from data with missing hits. Figure~\ref{fig:missingHits} demonstrates the quantum template matching algorithm's ability to successfully return possible hit-pattern templates for data with missing hits, with high probability. 

The quantum methodologies presented in this article not only adeptly manage incomplete data but also underline the durability and adaptability of quantum algorithms when faced with real-world, inconsistent datasets. Furthermore, our exploration of track encoding and utilising templates for various hit patterns in detectors provides deeper insights into the potential of quantum techniques in collider physics. These findings emphasise the immense promise of quantum computing in high-energy physics.

To conclude, while acknowledging the challenges inherent to charged track finding, our research underscores the pivotal role and promise of quantum algorithms in setting new benchmarks and advancing the task of charged-track finding in particle collision studies.

\vspace{0.5cm}
\noindent{\textit{\textbf{Acknowledgments} We thank Patrick Dunne, Wayne Luk, Mikael Mieskolainen, Mathieu Pellen, Zhiqiang Que and Andy Rose for valuable discussions. We acknowledge the use of IBM Quantum services for this work and to advanced services provided by the IBM Quantum Researchers Program. The views expressed are those of the authors, and do not reflect the official policy or position of IBM or the IBM Quantum team.}

\newpage
\bibliographystyle{ieeetr}
\bibliography{refs}{}

\end{document}